%Paper: astro-ph/9312034
%From: Jonathan Katz <katz@howdy.wustl.edu>
%Date: Wed, 15 Dec 1993 14:38:11 -0600
%Date (revised): Tue, 15 Mar 1994 09:45:21 -0600

\magnification=\magstep1
\baselineskip=20pt
\centerline{Low Frequency Spectra of Gamma-Ray Bursts}
\bigskip
\centerline{J. I. Katz}
\centerline{Department of Physics and McDonnell Center for the Space
Sciences}
\centerline{Washington University, St. Louis, Mo. 63130}
\centerline{I: katz@wuphys.wustl.edu}
\bigskip
\centerline{Abstract}
\medskip
Particles with energies below the mean energy $E_0$ in relativistic shocked
plasmas should assume an equilibrium energy distribution.  This leads to a
synchrotron spectrum $F_\nu \propto \nu^{1/3}$ up to approximately the
critical frequency $\nu_0$ of an electron with the energy $E_0$.
Application to GRBs implies that a burst with $10^{-5}$ erg/cm$^2$s of soft
gamma-rays and $h \nu_0 = 300$ KeV should be about 18th magnitude in visible
light and a few $\mu$Jy at 1 GHz (less if self-absorbed).
\vfil
\noindent
Subject headings: Gamma-rays: Bursts---Shock Waves---Acceleration of
Particles
\eject
\centerline{1. Introduction}
\medskip
A model of gamma-ray bursts (GRB) has been proposed in which a relativistic
fireball's debris shell (Shemi and Piran 1990) forms a relativistic
collisionless shock when it interacts with surrounding low-density gas (Rees
and M\'esz\'aros 1992, M\'esz\'aros and Rees 1993a, Katz 1994a).  The
shock-heated plasma radiates by the synchrotron process, producing the
observed gamma-rays.  It has been usual in astrophysics to assume that
shock-accelerated particles have a non-thermal distribution in energy.  In
this paper I argue that in the case of heating by a relativistic shock this
assumption is inapplicable to particles with energies below the mean particle
energy $E_0$, for which I predict a thermal equilibrium distribution function.
This implies a synchrotron spectrum $F_\nu \propto \nu^{1/3}$ below its peak.

Fermi (1949) derived the general result that the distribution function
produced by stochastic acceleration processes in a test-particle
approximation is proportional to a power of the particle energy, but a
dimensional argument is sufficient:  Any deviation from a power law would
define a characteristic energy, yet in these models there are no
characteristic energies between a particle's rest mass or injection energy
and an extremely high energy above which it is not magnetically confined to
the region of acceleration.  In the test-particle approximation the number
density $n$ of accelerated particles is indeterminate because there is an
essentially infinite reservoir of (nonrelativistic) thermal particles.  The
energy density $e$ of accelerated particles is then also indeterminate
because the acceleration is driven by a reservoir of fluid kinetic energy in
the shock or in plasma or hydrodynamic turbulence, an indeterminate
fraction of which is converted to particle acceleration.  The prediction of
a power-law differential distribution of particle energy under these
conditions
$$N(E) \propto E^{-p}, \eqno(1)$$
where $N(E)$ is the number of particles per unit energy per unit volume, is
borne out by observations of cosmic rays and of power-law synchrotron
spectra radiated by accelerated particles
$$F_\nu \propto \nu^{-s} \eqno(2)$$
in many astronomical objects; in optically thin synchrotron theory (Rybicki
and Lightman 1979) $s = (p - 1)/2$ if $p > 1/3$.  In order that the total
number and energy of accelerated particles be finite, a distribution of the
form (1) must have at least one characteristic energy $\sim E_0$ at which $p$
changes, with $p = p_< < 1$ for $E \ll E_0$ and $p = p_> > 2$ for $E \gg E_0$.
The corresponding synchrotron spectral indices $s_< < 0$ and $s_> > 1/2$.
\bigskip
\centerline{2. Relativistic Shocked Plasmas}
\medskip
It is often assumed that a power-law particle distribution function (Eq. 1)
will apply to any collisionless gas of shock-accelerated particles.
However, Fermi's test particle approximation which led to Eq. (1) is
inapplicable to relativistic collisionless shocks\footnote*{The argument
presented by Katz (1994a) for $p = 3/2$ and $s = 1/4$ is wrong;
nonrelativistic shock acceleration theory is inapplicable, and the assumed
compression ratio is inconsistent with the (correct) values cited elsewhere
in that paper.}.  In such a shock $n$ and $e$ are determined by the jump
conditions, and the mean energy per particle sets the natural energy scale
$E_0 \equiv e/n$.  At the low densities at which shocks may be considered
collisionless the rate of pair production is negligible.

In order for a shock to occur there must be an irreversible process, which
mixes the distribution function in phase space, increasing a coarse-grained
entropy.  No general theory of collisionless shocks exists.  The most
economical hypothesis is that made by Gibbs---that in the final state
all microstates consistent with the constraints (on $n$ and $e$) are equally
probable.  This assumption defines a microcanonical ensemble (Reif 1965),
and was also made by Lynden-Bell (1967) in his discussion of violent
gravitational relaxation.  The resulting distribution of particle energies,
assuming relativistic kinematics and nondegeneracy, is that of thermal
equilibrium
$$N(E) \propto E^2 \exp(-3E/E_0); \eqno(3)$$
the ``temperature'' $k_B T = E_0/3$, and the parameters $p_< = -2$ and $p_>
\to \infty$.  This argument for Eq. (3) depends essentially on the existence
of the constraints.

The relativistic Maxwellian (Eq. 3) was found by Gallant, {\it et al.}
(1992) in simulations of relativistic shocks in electron-positron plasmas.
In simulations of relativistic shocks in electron-positron-ion plasmas
Hoshino, {\it et al.} (1992) found an excess of energetic positrons with a
power-law distribution; only a fraction of the positron energy appeared in
the nonthermal particles, with most of it remaining in the thermal
population.  The simulations assumed greater magnetic energy density in the
upstream (unshocked) plasma than is likely to be found in GRB, and no
simulation appears to have addressed the problem of relativistic shocks in
electron-ion plasmas.  ``Plasma turbulent reactor'' theories (Norman and Ter
Haar 1975) are consistent with either Maxwellian or power-law distributions.

Sommer, {\it et al.} (1994) observed from one GRB a power-law gamma-ray
spectrum with $s \approx 1$ extending up to $h \nu \sim 1$ GeV.  This
radiation must be produced by electrons with $E > E_0$.  The observed $s$ is
consistent with the value $p = 3$ suggested by Norman and Ter~Haar (1975).
However, Hoshino, {\it et al.} (1992) found $p \approx 2$ in simulations
with one space and two momentum coordinates; it is unclear whether this
result can be applied directly to three dimensional plasmas, but the
increased dimensionality is more likely to decrease than to increase $p$.

This paper is concerned with radiation by particles for which $E \ll E_0$.
In this limit Eq. (3) becomes
$$N(E) \propto E^2, \eqno(4)$$
consistent with all the simulations (allowing for their reduced
dimensionality).

In a collisionless shock Eqs.~(3) or (4) must result from the interaction of
coarse-grained clumps in phase space, rather than from single-particle
collisions.  These clumps interact electromagnetically through plasma
turbulence, and we may regard the fields as the means by which the clumps
interact, in analogy with the gravitational interaction of coarse-grained
clumps considered by Lynden-Bell.

Order-of-magnitude arguments may permit an estimate of the magnitude of the
fields.  If the phase space clumps are sufficiently long-lived (as, for
example, may be the electrostatic clumps discussed by Dupree [1982]), they
may come to equilibrium with each other and with the turbulent fields,
considered as independent degrees of freedom.  This hypothesis is distinct
from that of violent relaxation.  The brightness temperature $T_b$ of the
turbulence is defined by
$$k_B T_b \equiv {{\cal F}_\nu c^2 \over \nu^2}, \eqno(5)$$
where ${\cal F}_\nu$ is the spectral density of the turbulence.  In a
relativistic plasma most modes will have relativistic phase and group
velocities, so we may approximate
$${\cal F}_\nu \approx {B^2 c \over 8 \pi \Delta \nu}, \eqno(6)$$
where $B$ is the turbulent magnetic field and $\Delta \nu$ is its spectral
bandwidth.  A clump of size $\lambda$ will contain $\approx n \lambda^3$
particles and a total energy $\approx n \lambda^3 E_0 \approx k_B T_c$,
where $T_c$ is the kinetic temperature of the clumps.  Taking $\Delta \nu
\approx \nu$ and $\lambda \approx c/\nu$ and equating $T_c$ and $T_b$ yields
$$e = n E_0 \approx {B^2 \over 8 \pi}. \eqno(7)$$
This demonstrates that equipartition between clumps and plasma turbulence is
consistent with the energetics, and lends credibility to the equipartition
of particle and magnetic energy assumed by Katz (1994a).

The length $\lambda$ is determined by the wavelengths of the fastest
growing plasma instabilities.  These may be two-stream instabilities when
the plasmas first interpenetrate, followed by the Weibel (1959) instability
(excitation of transverse electromagnetic waves by velocity space anisotropy).
For strong anisotropy the Weibel instability may saturate at
approximate magnetic equipartition.  Equation (7) also implies approximate
equality between the plasma and gyro-frequencies $\omega_p$ and $\omega_g$,
so that the distribution of $\lambda$ is probably peaked around $\lambda
\sim 2 \pi c / \omega_p \sim 2 \pi c / \omega_g$, with ${\cal F}_\nu$ peaked
around $\nu \sim c / \lambda$.  Fields of the magnitude given by Eq. (7)
imply an equipartition time $O(\omega_p^{-1}) \sim O(\omega_g^{-1})$, and
the lifetime of the clumps need only be of the same order.

Electron-ion equipartition cannot result from the very slow single-particle
interactions, and requires electromagnetic coupling between charge- or
current-unneutralized clumps in electron and ion phase space.  In a
relativistic plasma turbulence resembles propagating electromagnetic waves,
with electric fields ${\cal E} \sim B$.  Then the charge imbalance $\vert
n_i - n_e \vert \sim O(n)$, the electrostatic potentials are $O(E_0 /e)$,
and electron-ion equipartition is assured.  These estimates need to be
tested by plasma simulation.

Finally, a nonequilibrium distribution can also be shown to relax to the
equilibrium (Eq. 3) if an H-theorem is satisfied, which follows from the
assumption of molecular chaos (Liboff 1969).  This assumption holds if the
phase space clumps interact weakly, but is likely to be more
general---H-theorems are observed to hold in dense and strongly coupled
systems for which it is hard to justify the assumption of molecular chaos.
Strong two-particle (or two-clump) momentum correlations are disrupted, at
least in ensemble average, as the particles (or clumps) interact with many
others, and thermodynamic equilibrium is observed even though the assumption
of molecular chaos is hard to defend.
\bigskip
\centerline{3. Predicted Spectra}
\medskip
For relativistic electron distributions with $p < 1/3$ the synchrotron
spectrum at frequencies below the critical frequency is dominated by the
radiation of the most energetic electrons (Jackson 1975), and $s = - 1/3$
(rather than $(p - 1)/2$).  Thus the particle distribution (Eq. 4) leads to
a predicted radiation spectrum
$$F_\nu \propto \nu^{1/3}, \eqno(8)$$
which in GRB should extend roughly from X-rays to radio frequencies.
Distributions which differ from Eq.~(4) only for $E > E_0$, even if
non-thermal, also lead to Eq.~(8) for $\nu \ll \nu_0$, where $\nu_0$ is the
critical synchrotron frequency (Doppler-shifted to the observer's frame) for
electrons with the local $E_0$ in the local magnetic field.  This predicted
spectrum survives averaging over emission regions with a range of $E_0$,
field, and Doppler shift.

Schaefer (1994) found $N_\nu \propto \nu^{-0.7}$ in GRB at soft X-ray
energies, corresponding to $F_\nu \propto \nu^{0.3}$, consistent with
Eq.~(8).  The ``redder'' (larger $s$) spectra of GRB at harder X-ray and
gamma-ray energies reflect the radiation of particles with $E > E_0$, whose
distribution is no longer described by Eq. (4), producing a gradual roll-off
below Eq.~(8) in the integrated spectrum with increasing frequency.

Optically thin classical relativistic synchrotron radiation leads to the
very general prediction $s \ge - 1/3$.  Observation of $s < - 1/3$ in
GRB would imply absorption, which may be photoelectric (Schaefer 1993) in
soft X-rays, by dust or gas in visible and ultraviolet light, or
self-absorption (M\'esz\'aros and Rees 1993b, Paczy\'nski and Rhoads 1993,
Katz 1994ab).  Because the radiation at lower frequencies is dominated by
electrons with $E \approx E_0$, in the self-absorbed region $s = -2$ rather
than the value $s = - 5/2$ found (Rybicki and Lightman 1979) when $p > 1/3$.

The spectrum of GRB may be normalized to the observed soft gamma-ray fluxes.
The normalization is necessarily very approximate, because source regions
are likely heterogeneous and the transition between Eq.~(8) and its
cutoff at high photon energies may therefore extend over several
decades of the integrated spectrum, including the soft gamma-ray region.  An
intense GRB with a flux $10^{-5}$ erg/cm$^2$s in a bandwidth of 400 KeV
around $h\nu = 300$ KeV has a flux there of $\approx 10$ mJy.  Extrapolation
leads to a flux of $\approx 0.2$ mJy in visible light, ($\approx 18$th
magnitude) and to $\approx 2\ \mu$Jy at 1 GHz.  The effective flux of a
brief transient measured by a broad-band receiver may be further reduced by
plasma dispersion.  Because these fluxes are low (and, in addition,
self-absorbed at low frequencies) intergalactic dispersion (Ginzburg 1973,
Palmer 1993, Katz 1994ab) and optical counterparts to GRB will be difficult
to observe.

For an actual GRB it is more accurate to extrapolate downward from the
highest frequency $\nu_{max}$ to which Eq.~(8) is observed; $\nu_{max}$ is
the smallest $\nu_0$ contributing to the observed radiation.  This may lead
to substantially higher visible and radio fluxes if $h\nu_{max} \ll 300$ KeV
because $s > - 1/3$ between $\nu_{max}$ and soft gamma-ray frequencies.
In the data of Schaefer (1994) $\nu_{max}$ appears to be $\sim 10$ KeV.

After the initial gamma-ray transient, a relativistic fireball will continue
to expand and to produce radiation up to a critical frequency which
decreases with time as the blast wave degrades (Paczy\'nski and Rhoads
1993; M\'esz\'aros and Rees 1993b, 1994; Katz 1994ab).  In a uniform medium
(a very unrealistic assumption, as shown by the complex temporal structure of
observed GRB) the intensity at a given frequency will increase with time (in
a simple model $\propto t^{4/5}$) until the lowest critical synchrotron
frequency decreases to the frequency of observation; the peak visible
brightness is $\sim 50$ times brighter ($\approx 14$th magnitude) and follows
the $\gamma$-ray emission by $\sim 100$ times the latter's duration, while
at 1 GHz the peak flux is $\sim 5000$ times brighter ($\sim 10$ mJy) and lags
by $\sim 40,000$ times the $\gamma$-ray duration (Katz 1994ab).  However,
at all times the spectrum (Eq. 8) should be observed, allowing for
absorption, in the very broad range of frequencies between $\nu_{max}$ and
the onset of self-absorption.
\bigskip
\centerline{4. Discussion}
\medskip
This paper makes two predictions for the characteristic spectrum of
synchrotron radiation produced by relativistic shocks.  The first
prediction, $s_< < 0$, uses only the finiteness of the total particle number.
The second, more specific, prediction, $s_< = -1/3$ (Eq. 8), is based upon the
argument (\S 2) for $p_< = -2$ (Eq. 4), although any distribution with
$p < 1/3$ is sufficient to lead to $s = - 1/3$.

The arguments presented here for relativistic shocks are general, and not
specific to GRB.  For example, they should apply to relativistic blast wave
models of AGN such as those proposed by Blandford and McKee (1977).

I thank P. Diamond and B. E. Schaefer for discussions, an anonymous referee
for constructive criticism, and NASA NAGW-2918
for support.
\bigskip
\centerline{References}
\def\ref{\medskip \hangindent=20pt \hangafter=1}
\parindent=0pt
\ref
Blandford, R. D. \& McKee, C. F. 1977 MNRAS 180, 343
\ref
Dupree, T. H. 1982 P. Fluids 25, 277
\ref
Fermi, E. 1949 PRev 75, 1169
\ref
Gallant, Y. A. {\it et al.} 1992 ApJ 391, 73
\ref
Ginzburg, V. L. 1973 Nature 246, 415
\ref
Hoshino, M. {\it et al.} 1992 ApJ 390, 454
\ref
Jackson, J. D. 1975 Classical Electrodynamics (Wiley, New York)
\ref
Katz, J. I. 1994a ApJ 422, 248
\ref
Katz, J. I. 1994b Huntsville Gamma-Ray Burst Workshop, eds. G. Fishman, K.
Hurley, J. Brainerd (AIP, New York) in press
\ref
Liboff, R. L. 1969 Introduction to the Theory of Kinetic Equations (Wiley,
New York)
\ref
Lynden-Bell, D. 1967 MNRAS 136, 101
\ref
M\'esz\'aros, P. \& Rees, M. J. 1993a ApJ 405, 278
\ref
M\'esz\'aros, P. \& Rees, M. J. 1993b ApJ 418, L59
\ref
M\'esz\'aros, P. \& Rees, M. J. 1994 Huntsville Gamma-Ray Burst Workshop,
eds. G. Fishman, K. Hurley, J. Brainerd (AIP, New York) in press
\ref
Norman, C. A. \& Ter Haar, D. 1975 Phys Rep 17, 307
\ref
Paczy\'nski, B. \& Rhoads, J. E. 1993 ApJ 418, L5
\ref
Palmer, D. M. 1993 ApJ 417 L25
\ref
Rees, M. J. \& M\'esz\'aros, P. 1992 MNRAS 258, 41p
\ref
Reif, F. 1965 Fundamentals of Statistical and Thermal Physics (McGraw-Hill,
New York)
\ref
Rybicki, G. B. \& Lightman, A. P. 1979 Radiative Processes in Astrophysics
(Wiley, New York)
\ref
Schaefer, B. E. 1993 Compton Gamma-Ray Observatory, eds. M. Friedlander, N.
Gehrels, D. J. Macomb (AIP, New York) p. 803
\ref
Schaefer, B. E. 1994 Huntsville Gamma-Ray Burst Workshop, eds. G. Fishman,
K. Hurley, J. Brainerd (AIP, New York) in press
\ref
Shemi, A. \& Piran, T. 1990 ApJ 365, L55
\ref
Sommer, M. {\it et al.} 1994 ApJ 422, L63
\ref
Weibel, E. S. 1959 PRL 2, 83
\vfil
\eject
\bye
\end